\documentclass[%
preprint,
titlepage,
amsmath,amssymb,
aps,
onecolumn, 
superscriptaddress,
nofootinbib,
]{revtex4-2}

\usepackage{graphicx}
\usepackage{dcolumn}
\usepackage{bm}
\usepackage[utf8]{inputenc}

\usepackage{color}

\usepackage{float}

\usepackage{hyperref}

\newcommand{\alphas}{\alpha _{\rm s}}
\newcommand{\betas}{\beta _{\rm s}}
\newcommand{\Msun}{M_{\odot}}

\newcommand{\rhom}{\rho _{\rm m}}
\newcommand{\rhos}{\rho _{*}}
\newcommand{\mstar}{m_{*}}
\newcommand{\fb}{f_{\rm b}}

\usepackage{aas_macros}

\usepackage{xcolor}

\begin{document}


\title{Running Spectral Indices as a Solution \\to the JWST Early Galaxy Tension}
\title{Blue-tilted spectral running and the JWST early galaxy tension}

\author{Mikage U. Kobayashi}
\affiliation{Particle and Nuclear Physics Program, Graduate University for Advanced Studies (SOKENDAI), 1-1 Oho, Tsukuba, Ibaraki 305-0801, Japan}
\affiliation{Theory Center, IPNS, High Energy Accelerator Research Organization (KEK), 1-1 Oho, Tsukuba, Ibaraki 305-0801, Japan}
\author{Gen Chiaki}%
\affiliation{%
 National Institute of Technology (KOSEN), Kochi College \\
 200-1 Monobe, Nankoku, Kochi 783-8508, Japan 
}%
\author{Kazutaka Kimura}%
 \affiliation{Astronomical Institute, Graduate School of Science, Tohoku University, 6-3 Aramaki, Aoba, Sendai 980-8578, Japan}
\author{Kazuyuki Akitsu}%
\affiliation{Theory Center, IPNS, High Energy Accelerator Research Organization (KEK), 1-1 Oho, Tsukuba, Ibaraki 305-0801, Japan}
\author{Kazunori Kohri}
\affiliation{Division of Science, National Astronomical Observatory of Japan, and SOKENDAI, 2-21-1 Osawa, Mitaka, Tokyo 181-8588, Japan}
\affiliation{Department of Astronomy, The University of Tokyo, Bunkyo-ku, Hongo, Tokyo 113-0033, Japan}
\affiliation{Theory Center, IPNS, High Energy Accelerator Research Organization (KEK),
1-1 Oho, Tsukuba, Ibaraki 305-0801, Japan}
\affiliation{Kavli IPMU (WPI), UTIAS, The University of Tokyo, Kashiwa, Chiba 277-8583, Japan}
\author{Tomo Takahashi}%
 \affiliation{Department of Physics, Saga University, Saga 840-8502, Japan}
\author{Kazuyuki Omukai}%
 \affiliation{Astronomical Institute, Graduate School of Science, Tohoku University, 6-3 Aramaki, Aoba, Sendai 980-8578, Japan}

\date{\today}

\begin{abstract}
Recently, the James Webb Space Telescope (JWST) collaboration has found
the unexpectedly large abundance of massive galaxies 
with stellar masses of $\sim 10^{10}~M_{\odot}$ at high redshifts $z \simeq 6.5 - 9$ compared with the prediction of the standard $\Lambda$CDM model.
As a possible solution to the tension, we consider a blue-tilted spectrum of density perturbations with a positive running.
We find that, for $\alpha_s \simeq 0.02$ and $\beta_s \simeq 0.02$, a joint analysis with
CMB observations shows that the tension can be resolved at the 1$\sigma$ confidence level. 
Such a blue-tilted spectrum is also plausible from the perspective for formations of primordial black
holes on much smaller scales in the early Universe.
\end{abstract}

\maketitle


\section{\label{sec:intro}Introduction}

Recent observations with the James Webb Space Telescope (JWST) have uncovered a population of galaxies at very high redshifts ($z \simeq 6.5 – 9.0$) whose inferred stellar masses and number densities challenge expectations from the standard $\Lambda$CDM paradigm \cite{Labbe23,Boylan-Kolchin:2022kae,lovell2023extreme,Haslbauer:2022vnq}. Reconciling these findings within the conventional framework typically requires invoking extremely efficient and rapid star formation in early halos, in some cases pushing the limits of theoretical plausibility \cite{harikane2022comprehensive, Tacchella:2018qny, Behroozi:2020jhj}. This discrepancy has intensified efforts to understand the physical mechanisms driving the rapid assembly of the earliest galaxies.

One possible way to alleviate this tension is to modify the primordial curvature perturbations at small scales. In particular, scenarios in which the primordial power spectrum acquires a blue tilt at small scales can significantly enhance the abundance of early-forming low-mass halos, thereby allowing the observed galaxy population to be reproduced without requiring unrealistically high star formation efficiencies. Previous studies \cite{Parashari:2023cui,Hirano:2023auh,Colazo:2024jmz} have explored this possibility using simplified, piecewise parameterizations of a blue-tilted power spectrum around a pivot scale, which capture the essential phenomenology of such models.

In this work, we consider a blue-tilted primordial power spectrum characterized by a running spectral index, $\alpha_s$, and its running, $\beta_s$, which is physically motivated by some inflationary models~\cite{Lyth:1996kt,Leach:2000ea,Kohri:2007qn,Alabidi:2009bk,Drees:2011hb,Kobayashi:2010pz,Kobayashi:2012ba,Byrnes:2018txb,Furuta:2025fbh, Kumar:2025gon}. 
These parameters are constrained by Planck \citep{Planck18}, with $\alpha_s = 0.0011 \pm 0.0099$ and $\beta_s = 0.009 \pm 0.012$. In addition, we systematically explore the region of parameter space that can reproduce the galaxy population observed by JWST while remaining consistent with current observational constraints. In doing so, we evaluate whether such a physically motivated modification of the primordial power spectrum can provide a viable resolution to the tension between early galaxy formation and the standard $\Lambda$CDM paradigm.
Throughout this work, we use the cosmological parameters $\Omega _{\rm m} = 0.3099$, $\Omega _{\rm b} = 0.04868$, $\Omega _{\Lambda} = 0.6901$, and $h = 0.6781$ \citep{Planck18}.

\section{Method}

\subsection{\label{sec:level2}
Primordial power spectrum with runnings and halo mass function
}

In this subsection, 
we describe how the primordial power spectrum is
characterized using spectral indices and their runnings and its impact on the halo mass function (HMF). 
Inflationary models that 
could produce a blue-tilted  running spectral-index
within slow-roll approximations include, for example, 
mutated hybrid inflation models~\cite{Lyth:1996kt}, type-III hilltop inflation models~\cite{Leach:2000ea,Kohri:2007gq}, running-mass inflaton models \cite{Drees:2011hb}, modulated inflaton potential \cite{Kobayashi:2010pz}, some classes of curvaton models \cite{Kobayashi:2012ba}, ultra-slow-roll inflation~\cite{Byrnes:2018txb}, tachyonic trap during inflation~\cite{Furuta:2025fbh}, etc. Below we do not specify the mechanism by which large spectral runnings can be realized, but instead focus phenomenologically on its impact on the halo mass function.

The power spectrum of curvature perturbations with running spectral indices is given by
\begin{equation}
    \begin{aligned}
        P_{\zeta}(k) = A_s (k_p) \left(\frac{k}{k_p}\right)^{n_s -1 + \frac{1}{2} \alpha_s \ln\left(\frac{k}{k_p}\right) + \frac{1}{6}\beta_s \ln^2\left(\frac{k}{k_p}\right)} ,
    \end{aligned}
\end{equation}
where 
the spectral index is expanded up to second order in $\ln(k/k_p)$.
$A_s$ is the amplitude of the primordial curvature power spectrum, $n_s$ is the spectral index, $k_p$ is the pivot scale, $\alpha_s$ is the running of the spectral index, and $\beta_s$ is the running of the running of the spectral index.
In our analysis, we adopt the pivot scale $ k_p = 0.05 \mathrm{Mpc}^{-1}$ and fix the spectral index and the amplitude as 
$n_s = 0.96$,  $A_s = 2.100549 \times 10^{-9}$ as reported by the Planck Collaboration 2018~\citep{Planck18}.
We verified that adopting the ACT DR6 value ($n_s = 0.97$)\cite{AtacamaCosmologyTelescope:2025blo}, has a negligible impact on the results over the parameter range considered in this work.

The matter power spectrum is obtained from the primordial curvature power spectrum including the running spectral index $\alpha_s$ and its running $\beta_s$.
We compute the linear matter power spectrum using CAMB~\cite{Lewis:1999bs, Lewis:2002ah}, which incorporates the effects of the transfer function and the linear growth of perturbations.

To connect the matter power spectrum to halo abundances, we employ the Sheth-Tormen (ST) mass function \cite{Gavas:2022iqb}, given by
\begin{equation}
\begin{aligned}
f(\nu)=A(p) \nu \sqrt{\frac{2q}{\pi}} \left[1 + (q \nu^2)^{-p}\right]\exp\left(-\frac{q\nu^2}{2}\right),
\end{aligned}
\end{equation}
where $\nu = \delta_c / (D_+(z)\sigma(m))$ with a redshift $z$ and halo mass $m$ , and $D_+(z)$ denotes the linear growth factor, while $\delta_c = 1.69$ is the critical overdensity for spherical collapse.
The mass variance $\sigma^2(m)$ is defined as
\begin{equation}
\begin{aligned}
\sigma^2(m) = \int^{\infty}_{0} \frac{\mathrm{d}k}{k} \frac{k^3 P(k)}{2\pi^2} W^2(k,m),
\end{aligned}
\end{equation}
where $P(k)$ is the matter power spectrum and $W(k,m)$ is the window function.
The matter power spectrum can be decomposed into a factor evaluated without the blue tilt and a factor with the blue tilt, as
$P(k) = P(k)|_{\alpha_s,\beta_s=0}P_{\rm tilt}(k,\alpha_s,\beta_s)$ since the matter power spectrum is proportional to the primordial power spectrum.
Throughout this work, we employ a Gaussian window function, given by
\begin{equation}
    \begin{aligned}
        W(k,m) = \exp\left(-\frac{1}{2}(kR)^2\right)
    \end{aligned}
\end{equation}
where $R^2$ is the variance. 
The normalization factor $A(p)$ is given by
\begin{equation}
\begin{aligned}
A(p) = \left[1 + \frac{2^{-p}\Gamma(0.5 - p)}{ \sqrt{\pi}}\right]^{-1}.
\end{aligned}
\end{equation}
In this paper, we adopt $p=0.3$ and $q=0.75$\cite{Sheth:1999su, Sheth:2001dp}.

Finally, the comoving number density of halos with mass $m$ is expressed as
\begin{equation}
\begin{aligned}
\frac{\mathrm{d}n}{\mathrm{d}\ln m} = \frac{\bar\rho}{m}\frac{\mathrm{d}\ln \sigma^{-1}}{\mathrm{d}\ln m} f(\nu),
\end{aligned}
\label{eq:HMF}
\end{equation}
where $\bar\rho$ denotes the mean matter density of the Universe.

This framework allows us to quantify the impact of the running of the primordial power spectrum on the abundance of dark matter halos.

\subsection{Cumulative stellar mass function}
Following the procedure described in the previous subsection, we can calculate the mass function of DM halos for a given set of parameters, including the running parameters $\alpha_s$ and $\beta_s$.
Since DM halos are not directly observable, 
we need to convert the halo mass into the stellar mass, $\mstar$, of galaxies hosted by the halos to compare the model predictions with observations.
The conversion consists of two steps: (i) from DM halo mass to baryonic mass, and (ii) from baryonic mass to stellar mass.
The baryonic mass can be obtained from the DM halo mass using the baryon fraction, which is assumed to be the cosmic mean value, $\fb = \Omega _{\rm b} / \Omega _{\rm m} = 0.157$, as suggested by previous simulations \cite{Abel:2002}.
The stellar mass is then given by $\mstar = f _* \fb m$, where $f _*$ is the fraction of baryons converted into stars, i.e., the star formation efficiency (SFE).

Following Boylan-Kolchin \cite{Boylan-Kolchin:2022kae} by changing
$f_*$\cite{Parashari:2023cui, Hirano:2023auh,Colazo:2024jmz}, 
we estimate the upper bound on the cumulative stellar mass density (CSMD), 
above which the required stellar mass would exceed the baryonic mass available within the host halos, as
\begin{equation}
\rhos (> \mstar) = f _* \fb \rhom \left(> \frac{\mstar }{ \fb f _*} \right),
\end{equation}
where $\rhom$ is the cumulative halo mass density
given by
\begin{equation}
\rhom (>m) = \int _{m} ^{\infty} d\bar{m}~\bar{m} \frac{dn}{d\bar{m}}.
\end{equation}
$\sigma^2$ is proportional to the integral of $P(k)W(k,m)^2$ over $k$, and if $W(k,m)$ is sufficiently sharp, the contribution of the blue tilt to $\sigma^2$ can be approximately factored out.
Since the CSMD is proportional to the cumulative halo mass density, which in turn is proportional to the halo mass function, and the halo mass function is proportional to $\mathrm{d}\ln\sigma^{-1}/\mathrm{d}\ln m$, the contribution of the blue tilt can therefore be approximately factored out of the CSMD as well.

By comparing the CSMD with JWST observations, we constrain the running parameters, $\alphas$ and $\betas$, assuming SFEs in the range $0.20 \leq f_* \leq 0.30$ with an interval of $\Delta f_* = 0.01$ at redshifts of $z=8$ and $9$. 
The adopted range of $f_*$ is motivated by previous numerical studies of star cluster formation at high redshifts \cite{Kim:2018, Fukushima:2021, Krumholz:2019, Chon2024}. 
The running parameters $\alpha_s$ and $\beta_s$ are varied in the ranges of $-0.05 \leq \alphas \leq 0.05$ and $-0.05 \leq \betas \leq 0.05$ with an interval of $\Delta \alphas = \Delta \betas = 0.001$.

\subsection{JWST galaxy sample}
To compare our model with observations, we use the results reported by \citet{Labbe23}, which have posed a challenge to standard galaxy-formation models in the $\Lambda$CDM framework.
They identified 13 galaxy candidates at redshifts $z \sim 6.5$--9.0, based on JWST/NIRCam photometry, using color criteria designed to identify a Lyman break consistent with $z \gtrsim 7$.
Their redshifts and stellar masses were estimated through spectral energy distribution fitting, and the resulting values are summarized in Table 2 of \citet{Labbe23}.
\par
Using the redshifts and stellar masses of the observed galaxies, we compute the CSMD in each redshift bin as
\begin{equation}
    \rho_{*,{\rm obs}}(>m_*)
    =
    \frac{1}{V_{\rm obs}}
    \sum_i M_{*,i}
    \Theta(M_{*,i}-m_*),
\end{equation}
where $M_{*,i}$ is the stellar mass of the $i$-th galaxy, $V_{\rm obs}$ is the comoving survey volume, $\Theta$ is the Heaviside step function. The summation is taken over all galaxies in the corresponding redshift bin.  
Following \citet{Labbe23}, we adopt two redshift bins, $7<z<8.5$ and $8.5<z<10$, and compare them with the model predictions at $z=8$ and $z=9$, respectively.  
In each redshift bin, we estimate the CSMD, using the three most massive galaxies.
In principle, the uncertainties include both Poisson fluctuations and cosmic variance. 
However, because the Poisson uncertainty dominates in this case \citep{Labbe23}, we only consider the Poisson error, computed assuming Poisson
statistics following \citet{Gehrels86}.

\section{Result}

\begin{figure*}
\includegraphics[width=\textwidth]{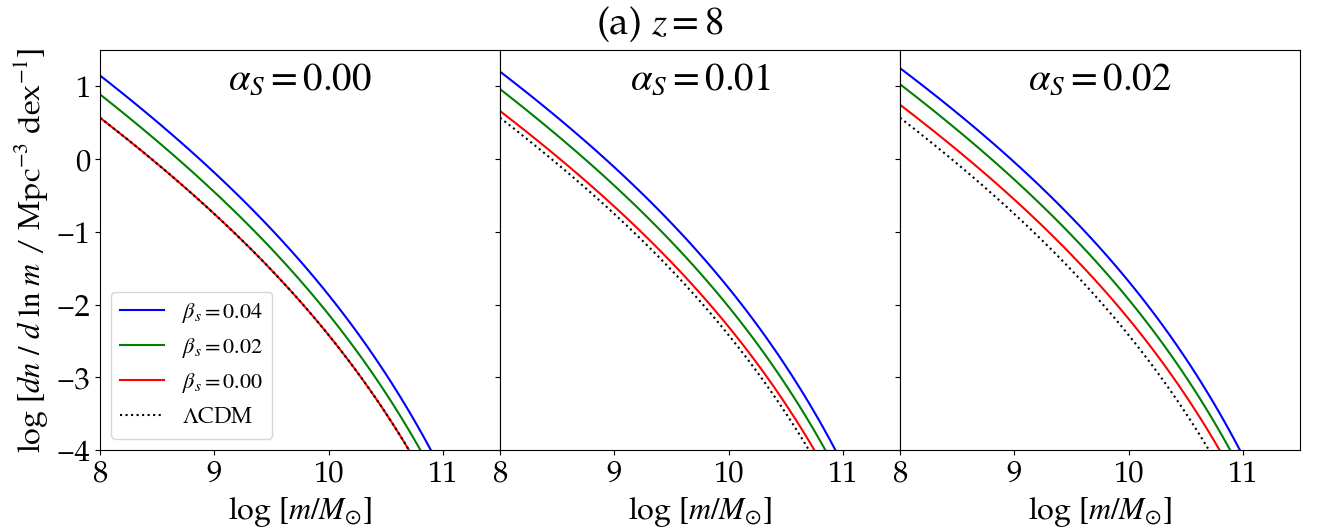}
\includegraphics[width=\textwidth]{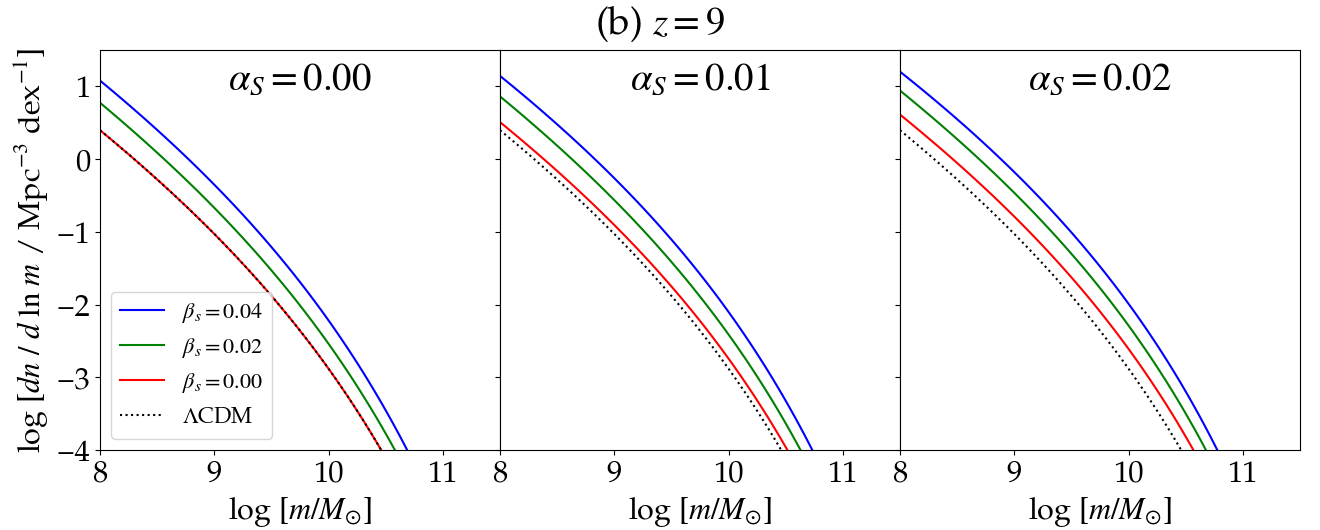}
\caption{\label{fig:HMF} Halo mass functions (HMFs) with $\alphas = 0.00$, $0.01$, and $0.02$ from left to right at (a) $z=8$ and (b) $z=9$.
The red, green, and blue curves in each panel show the results for $\betas = 0.00$, $0.02$, and $0.04$, respectively.
The black dotted curves show the result for the standard $\Lambda$CDM model ($\alphas = 0$ and $\betas = 0$).}
\end{figure*}

\begin{figure*}
\includegraphics[width=\textwidth]{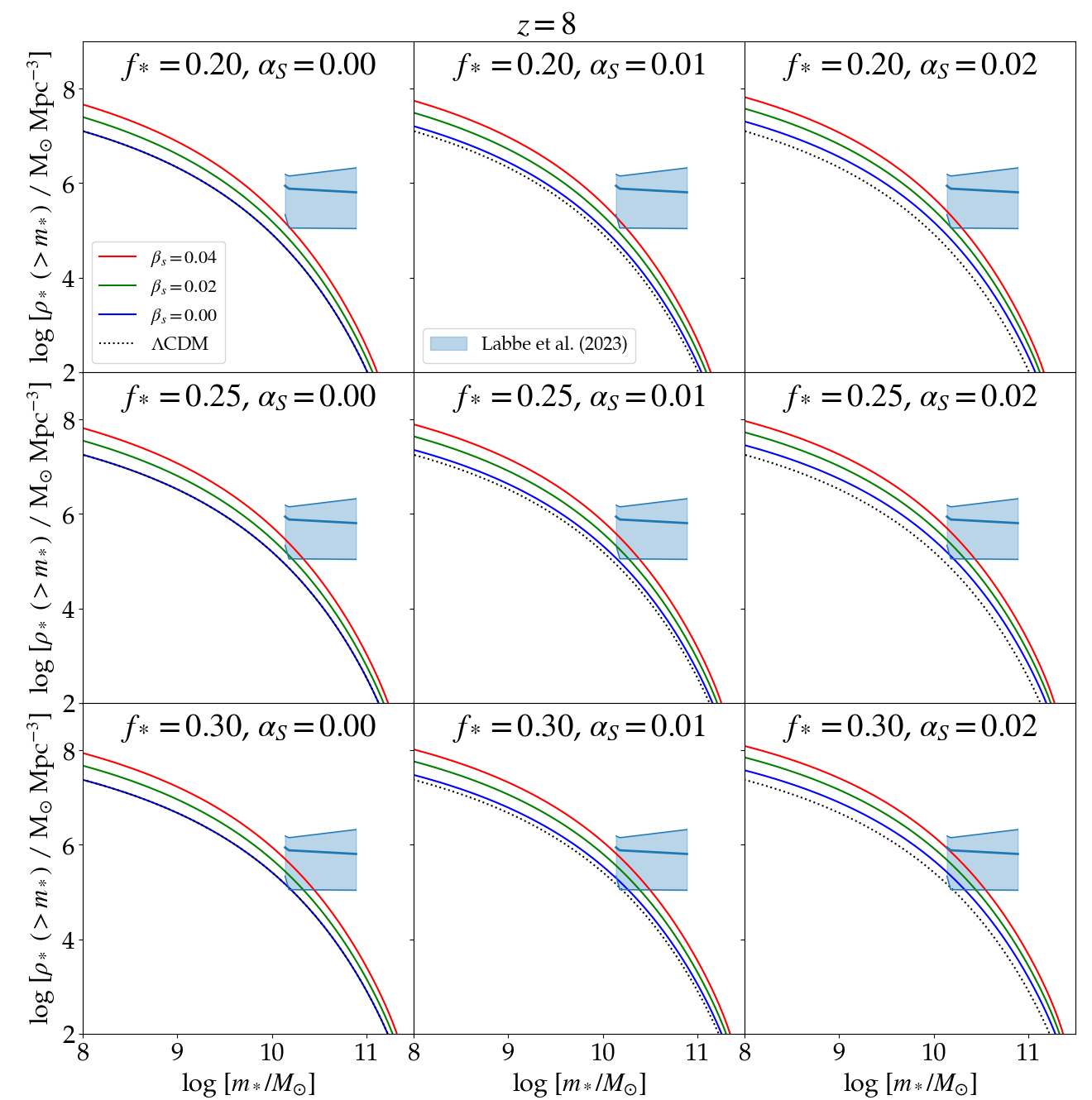}
\caption{\label{fig:csmd_z8} Cumulative stellar mass densities (CSMDs) at redshift $z=8$.
The top, middle, and bottom rows show the results for star formation efficiencies of $f_* = 0.20$, $0.25$, and $0.30$, respectively.
The left, middle, and right columns show the results for $\alphas = 0.00$, $0.01$, and $0.02$, respectively.
The red, green, and blue curves in each panel show the results for $\betas = 0.00$, $0.02$, and $0.04$, respectively.
The black dotted curve show the result for the standard $\Lambda$CDM model ($\alphas = 0$ and $\betas = 0$).
The blue shaded region represents the CSMD of high-redshift galaxies observed by JWST \citep{Labbe23}.
}
\end{figure*}

\begin{figure*}
\includegraphics[width=\textwidth]{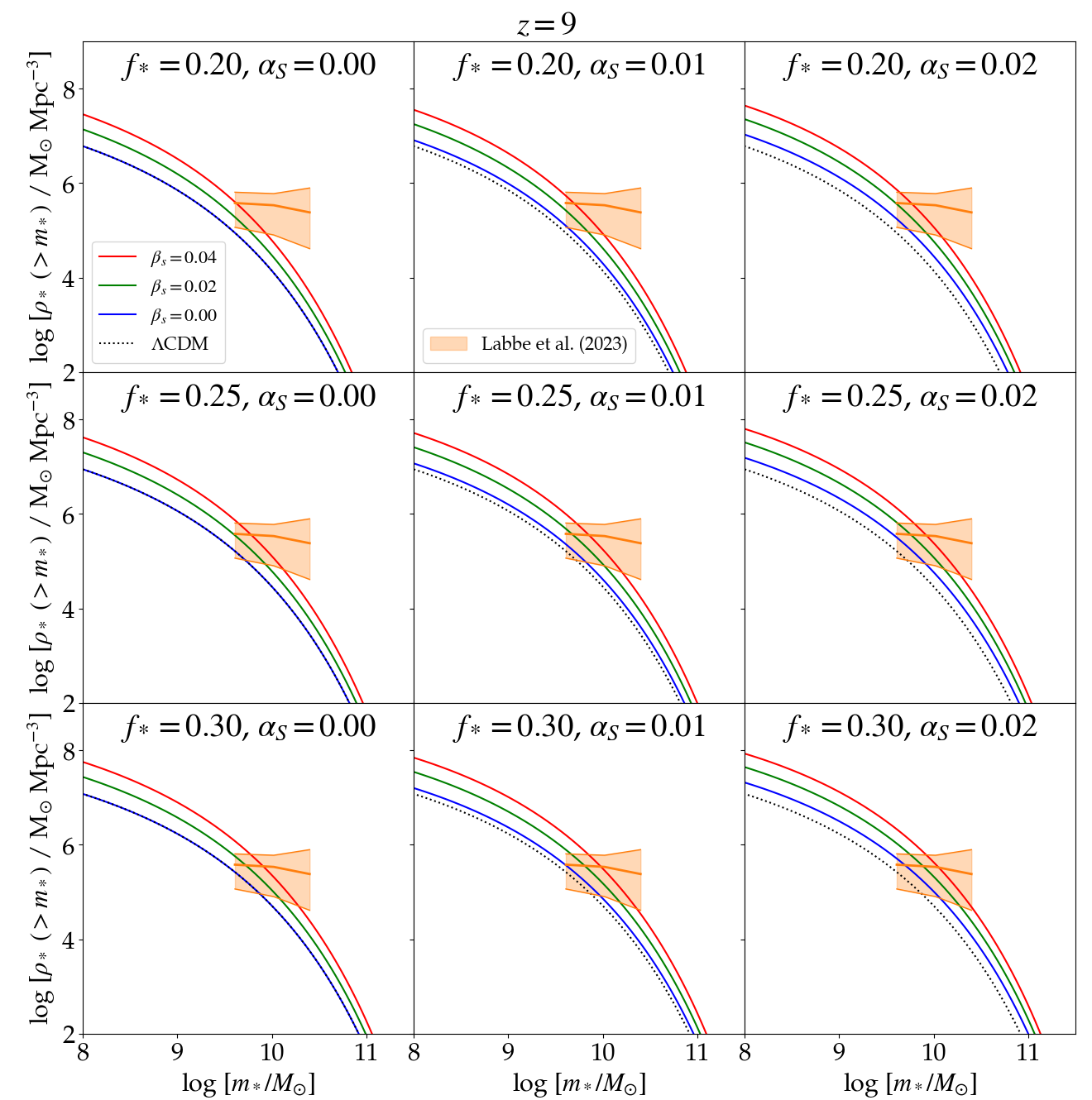}
\caption{\label{fig:csmd_z9} Same as Fig. \ref{fig:csmd_z8} but at  redshift $z=9$.}
\end{figure*}

\subsection{Halo mass function at high redshifts}

From Eq.~(\ref{eq:HMF}), we obtain the HMFs at redshifts $8$ and $9$ as shown in Fig.~\ref{fig:HMF} (a) and (b), respectively.
At each redshift, the left, middle, and right panels show the results for $\alphas = 0.00$, $0.01$, and $0.02$, respectively.
The blue, green, and red curves show the results for $\betas = 0.00$, $0.02$, and $0.04$, respectively.
In general, the halo abundance decreases with increasing halo mass, $m$.
The HMF follows a power-law distribution with a negative slope and exhibits an exponential cut-off at $m\gtrsim 10^{10}~\Msun$.
This indicates that massive galaxies hosted by such massive halos are intrinsically rare.
With increasing $\alphas$ and $\betas$, the CSMD also increases due to the enhancement of small-scale power by the running effect, and can exceed that predicted by the standard $\Lambda$CDM model ($\alphas = 0$ and $\betas = 0$; black dotted curves).

\subsection{Cumulative stellar mass density}

We calculate the upper bound of the CSMD at redshifts $z=8$ and $9$ as shown in Figs. \ref{fig:csmd_z8} and \ref{fig:csmd_z9}, respectively.
The top, middle, and bottom panels show the results for star formation efficiencies $f_* = 0.20$, $0.25$, and $0.30$, respectively.
The left, middle, and right panels show the results for $\alphas = 0.00$, $0.01$, and $0.02$, respectively.
The blue, green, and red curves show the results for $\betas = 0.00$, $0.02$, and $0.04$, respectively.
The CSMD is larger at lower redshift ($z=8$) because halos grow in mass through mergers.
We also find that the CSMD increases with increasing $f_*$ for fixed $\alphas$ and $\betas$, because the stellar mass associated with each halo becomes larger.
As expected, 
when $\alphas >0$ and $\betas >0$, the CSMD exceed that predicted by the standard $\Lambda$CDM model without runnings (black dotted curves), thereby promoting the formation of massive, bright galaxies at very high redshifts.

\subsection{Comparison with JWST data}

In Figs. \ref{fig:csmd_z8} and \ref{fig:csmd_z9}, we overplot the CSMD inferred from the observed abundance of high-redshift galaxies, shown by the blue and orange shaded regions, respectively.
Within each shaded region, the thick curve in the middle indicates the average value, while the upper and lower curves indicate the upper and lower bounds of the 1-$\sigma$ uncertainty.
The standard $\Lambda$CDM model with $\alphas = 0$ and $\betas = 0$ fails to explain the observed abundance of bright galaxies at the 1-$\sigma$ confidence level for $f_* = 0.20$ at both redshifts (see the black dotted curves in the top panels of Figs. \ref{fig:csmd_z8} and \ref{fig:csmd_z9}).
The standard $\Lambda$CDM model is marginally consistent with the observations $f_* \gtrsim 0.25$ (0.20) at $z=8$ (9, respectively) although the predicted CSMD is still below the observational average.

By introducing the running effect, the discrepancy at $z=8$ for $f_* = 0.20$ can be alleviated.
With $\alphas = 0.01$, the model falls within the 1-$\sigma$ uncertainlty for $\betas \gtrsim 0.04$.
For $\alphas = 0.02$, the observations can be explained with $\betas \gtrsim 0.02$.
When a slightly larger SFE, $f_* = 0.25$,  is assumed,
the CSMD predicted by the model
falls within the 1-$\sigma$ observational uncertainty for $\betas \gtrsim 0.02$ even with $\alphas = 0$.
In any case, the observed CSMD can be reproduced for all values of $f_*$ considered in our analysis when the running parameters satisfy $\alphas \geq 0.01$ and $\betas \geq 0.02$.
For $\alphas = 0.02$ and $\betas = 0.04$, the model reaches the observational average at $\mstar \sim 10^{10}~\Msun$.

At $z=9$, models with $\betas \gtrsim 0.02$ can reproduce the observations 
even for $\alphas = 0$.
For $\alphas \geq 0.01$, the predicted CSMD falls within the 1-$\sigma$ uncertainty even for $\betas=0$.
In particular, for $\alphas =0.02$, the model falls within the observatinal uncertainty over the stellar mass range $10 \lesssim \log \mstar \lesssim 10.5$.
However, we note that even with such large running parameters, we still cannot explain the high abundance of bright galaxies at $\log \mstar \gtrsim 10.5$.



\begin{figure*}
\includegraphics[width=\textwidth]{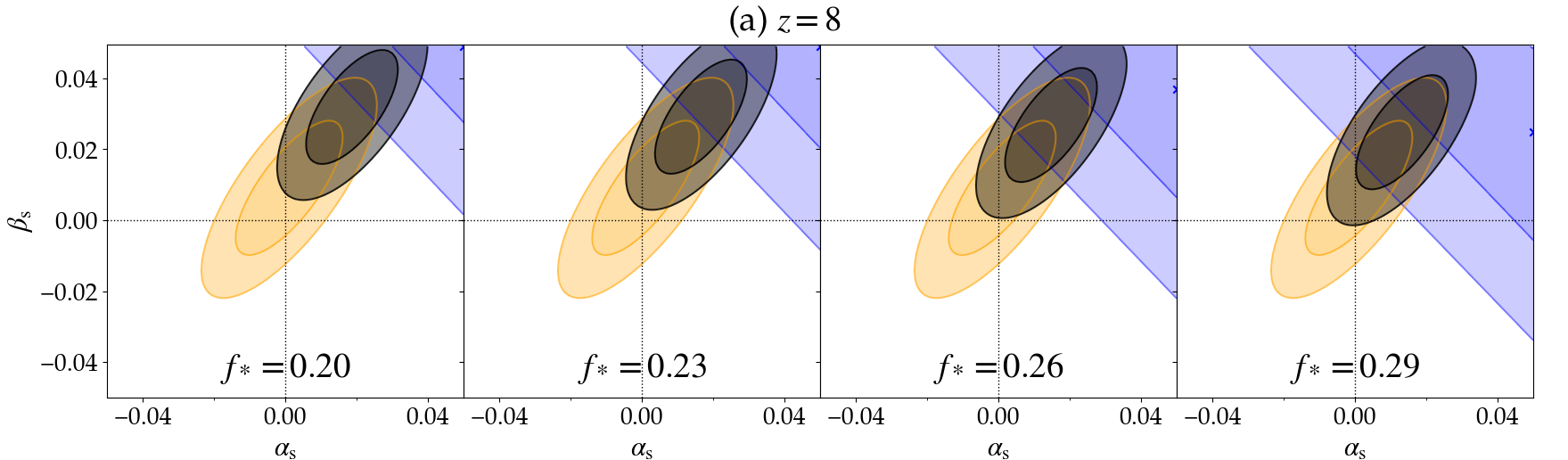} 
\includegraphics[width=\textwidth]{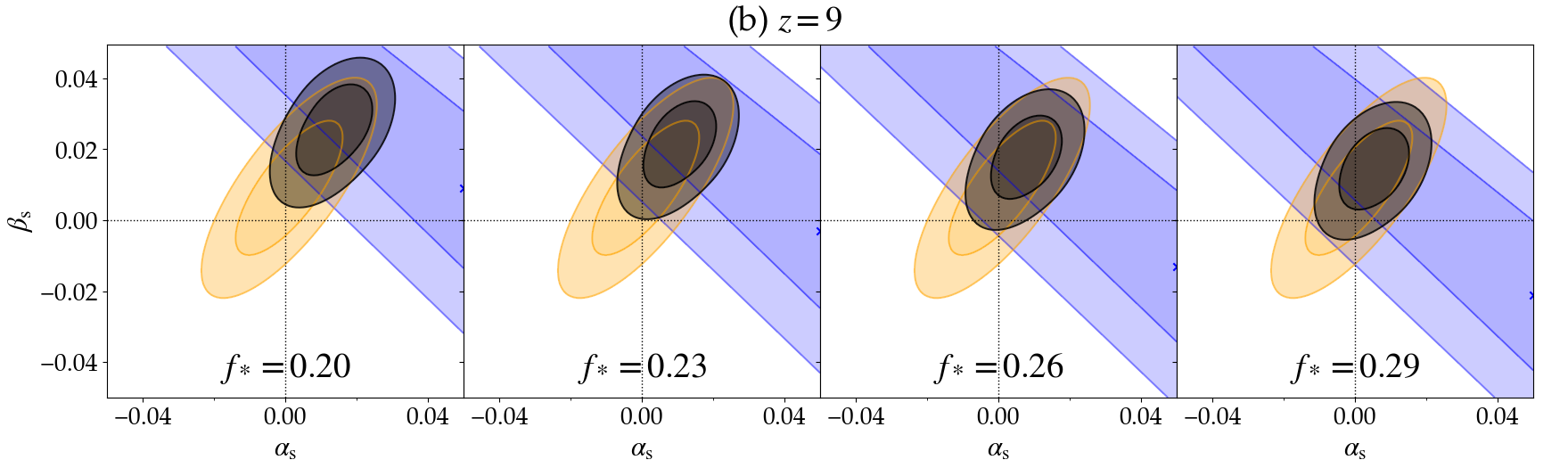} 
\caption{\label{fig:alpha_beta} Maginalized 1- and 2-$\sigma$ constraint contours for the running parameters $\alphas$ and $\betas$.
The blue and orange contours show the constraints obtained in this work and from Planck \citep{Planck18}, respectively, while the black shaded regions indicate the allowed parameter space from the combined JWST and Planck data.
The left, middle, and right panels correspond to $f_* = 0.20$, $0.23$, $0.26$, and $0.29$, respectively, at (a) $z=8$ and (b) $z=9$.}
\end{figure*}

\subsection{Constraint on running parameters}

By comparing the model predictions with observations, we constrain the running parameters $\alphas$ and $\betas$.
The blue shaded regions in Fig. \ref{fig:alpha_beta} show the constraint contours for $f_* = 0.20$, $0.23$, $0.26$, and $0.29$ from left to right panels at (a) $z=8$ and (b) $z=9$.
At $z=8$, for $0.20 \leq f_* < 0.30$, the hypothesis that both $\alphas$ and $\betas$ are zero
is rejected at more than the 2-$\sigma$ confidence level (CL) by the JWST observations.
This is because the observed CSMD is overall higher than the CSMD predicted by the standard $\Lambda$CDM model without runnings (see the black dotted curves in Fig. \ref{fig:csmd_z8}).
At $z=9$, 
the JWST data disfavours no runnings ($\alphas=0$ and $\betas=0$) at the $2$-$\sigma$ CL for $f_* = 0.25$ and at the $1$-$\sigma$ CL for $f_* = 0.30$.
The CL is smaller than at $z=8$ because the observed CSMD is closer to 
the no-running case
(see the black dotted curves in Fig. \ref{fig:csmd_z9}).

The running parameters have also been independently constrained by CMB measurements.
The constraint contour derived from the 2018 Planck collaboration \cite{Planck18} result is shown by the orange contour in Fig. \ref{fig:alpha_beta}, which indicates that
the no-running case is
favoured within the $1$-$\sigma$ CL.
However, when combined with our JWST constraints, the 
resulting constraint contours (black shaded regions in Fig. \ref{fig:alpha_beta}) become more distant from $\alphas = \betas = 0$ for smaller values of $f_*$.
Calculating the CSMD with an interval of $\Delta f_* = 0.01$, we find that the contours still deviate from 
the no-running case
for $f_* = 0.26$ at $z=8$ and $f_* = 0.23$ at $z=9$, even at the $2$-$\sigma$ CL.


\section{Discussion and Conclusion}

We have investigated the possibility that primordial density perturbations on small scales are enhanced relative to the standard $\Lambda$CDM model 
by considering a blue-tilted primordial power spectrum. 
This may provide a solution to the tension raised by JWST observations of overmassive galaxies at high-redshifts ($z \simeq 6.5 – 9.0$), for which the estimated cumulative stellar-mass densities are significantly larger than those predicted by the standard $\Lambda$CDM model.

In the standard $\Lambda$CDM model, we found that the formation of such massive galaxies at high redshifts is barely consistent with the JWST+Planck data only when $f_* > 0.26$ at $z=8$ and $f_* > 0.23$ at $z=9$
although the predicted CSMD still remains below the observational mean.
However, such high SFEs are unlikely in the local universe, where the SFE is typically only a few \% \cite{Biegel:2008, Kennicutt:2012}.
Even at high redshifts ($z \gtrsim 6$),  
the preferred values of $f_*$ is somewhat higher
than those predicted by cosmological simulations, which typically find $f_* \simeq 0.2$--$0.3$ \cite{Kim:2018, Fukushima:2021, Krumholz:2019}. 

In particular, in inflationary models with a blue-tilted spectrum, positive non-zero values of the 
running spectral index $\alpha_s$ and the running-of-running spectral index $\beta_s$ enhance primordial
density perturbations on small scales relative to
the standard $\Lambda$CDM model.
In this case, the abundance of low-mass halos formed in the early Universe increases, leading to higher number of galaxies at high redshifts. In fact, a joint analysis combining the JWST data with the Planck 2018 CMB constraints shows that
tension could be resolved at the 1-$\sigma$ confidence level for $\alpha_s \simeq 0.02$ and $\beta_s \simeq 0.02$. 

Such a blue-tilted spectrum can further produce larger density fluctuations, $\delta \sim {\cal O}(1)$, on much smaller scales in the range of
$k=10^6$~Mpc$^{-1}$ -- $10^{15}$~Mpc$^{-1}$, 
where 
primordial black holes (PBHs) may form. The PBHs have attracted significant attention as candidates for dark matter in terms of non-detections of the Hawking evaporation~\cite{Carr:2020gox,Coogan:2020tuf,Berteaud:2022tws} or the gravitational lensing~\cite{Niikura:2017zjd,Smyth:2019whb,Sugiyama:2026kpv}, as possible origins of the binary black hole merger events detected by LIGO-Virgo-KAGRA~\cite{Sasaki:2016jop,Franciolini:2021tla,Andres-Carcasona:2024wqk},
and as seeds of supermassive black holes~\cite{Kawasaki:2012kn,Kohri:2014lza,Kobayashi:2025jkg}.
Therefore, a blue-tilted spectrum is also highly promising from the perspective of the PBH formation.

Interestingly, even if the enhancement of primordial fluctuations is insufficient for efficient PBH formation, it may still significantly accelerate the formation of rare early halos or ultracompact minihalos~\cite{Kohri:2014lza,Nakama:2017qac,Inman:2022uvy}. In such cases, star formation could occur at extreme redshifts of $z \gtrsim 500$, where the high CMB temperature suppresses $\mathrm{H}_2$ cooling and induces nearly isothermal collapse via atomic cooling, resulting in the formation of supermassive stars with masses of order $10^5\,M_\odot$ \citep{Ito2024,Qin2025}. These objects would likely collapse into massive black holes, providing another possible route to massive black-hole seed formation in the very early Universe.

\begin{acknowledgments}
This work was in part supported by JSPS KAKENHI Grants
Nos. JP23KF0289, JP24K07027 (K. Kohri), JP25K01004 (T. Takahashi), JP24KJ0015 (K. Kimura) and MEXT KAKENHI Grants Nos. JP24H01825 (K. Kohri),  JP23H04515 (T. Takahashi) , JP25H01543 (T. Takahashi) , JP22H00149 (K. Omukai) and JP26H02060 (K. Omukai).
\end{acknowledgments}

\appendix



\nocite{*}

\bibliography{main}

\end{document}